\documentclass{article}
\usepackage{graphicx}
\usepackage[top=3cm, bottom=3cm, left=3.5cm, right=3.5cm]{geometry}
\usepackage{authblk}
\usepackage{dirtree}
\usepackage{booktabs}
\usepackage{tabularx}
\usepackage[superscript]{cite}
\usepackage{xurl}

\title{Dataset on residential electricity load profiles in Switzerland}
\author[1,4]{Katharina Kaiser}
\author[2]{Markus Kreft}
\author[3]{Marina González Vayá}
\author[1]{Gabriela Hug}
\affil[1]{Power Systems Laboratory, ETH Zurich, Switzerland}
\affil[2]{Chair of Information Management, ETH Zurich, Switzerland}
\affil[3]{Elektrizitätswerke des Kantons Zürich (EKZ), Switzerland}
\affil[4]{Corresponding author: kkaiser@ethz.ch}
\date{}

\begin{document}
\maketitle

\section*{Abstract}
Residential electricity profiles are undergoing significant changes due to the increasing adoption of distributed renewable generation, heat pumps, and electric vehicles. Up-to-date real-world measurements are essential for characterizing these evolving consumption patterns and providing a timely empirical foundation for research. This paper presents a dataset of 15-minute smart meter measurements from 2,447 residential installations in Switzerland, covering the years 2023 and 2024. The dataset encompasses both active and reactive energy (withdrawal and injection) of diverse installation types, including apartments, single-family houses, heat pumps, and common building areas. A subset of the associated customers participated in a pilot project featuring novel grid tariffs and automated load control. The corresponding price profiles and metadata are also included. The dataset aims to support research focusing on distribution grid modeling, load forecasting, and the assessment of time-varying tariffs.

\section{Background \& Summary}
The residential sector in Switzerland accounts for approximately one third of the country's annual electricity demand \cite{BfeSchweizerischeElektrizitätsstatistik2024} and plays an important role in the ongoing energy transition. Increasing shares of distributed renewable generation, heat pumps, and electric vehicles (EVs) are key components to reach national emission goals \cite{BFEEnergieperspektiven2050}{}. However, they also alter households' electricity profiles, posing challenges for the planning and operation of distribution grids and potentially leading to significant grid reinforcement needs \cite{WillemsenAuswirkungenStarken2022}{}.

Real-world electricity measurements can support research addressing these challenges in several ways. The data may be leveraged for simulations, the development of forecasting tools, analyses of the main contributors to grid-level peaks, or the assessment of measures such as time-varying price signals. Various overviews on open electricity datasets exist, both in research papers \cite{HabenReviewLow2021} and online \cite{ResourcesSIGENERGY}{}. The datasets vary in terms of time resolution, measurement level (e.g., device- or household-level), measured values (e.g., power only or also voltage and current), and provided metadata. Datasets in Switzerland include the ECO dataset \cite{BeckelECOData2014} with device-level and aggregate measurements of six households in 2012/13, a dataset on voltage and power measurements of 24 secondary substations and low-voltage cabinets in 2018/19 \cite{NespoliHierarchicalDemand2020}{}, and the HEAPO dataset \cite{BrudermuellerHEAPOOpen2025} with 15-minute smart meter measurements of 1,408 households with heat pumps between 2018 and 2024. Additionally, the utility CKW continuously publishes 15-minute consumption data of more than 100,000 private households and small businesses with an annual consumption of up to 25~MWh \cite{OpenDataCKW}{}. However, it does not include any metadata.

In contrast, this work introduces a dataset consisting of 15-minute-resolution smart meter data from 2,447 residential installations in the canton of Zurich, Switzerland, including metadata on the type of installation and the major loads connected. The measurements, recorded across 2023 and 2024, cover various types of installations, including apartments, single-family houses, heat pumps, and common installations in apartment buildings. Beyond the active energy withdrawal provided in the HEAPO and CKW datasets \cite{BrudermuellerHEAPOOpen2025, OpenDataCKW}{}, this dataset also includes the active energy fed back into the grid and the reactive energy at 15-minute resolution. A subgroup of 606 meters is linked to the pilot project OrtsNetz \cite{KaiserOrtsNetz2025}{}, which ran from October 2023 to December 2024 and featured experimental energy and grid tariffs. For a subset of customers who had agreed to the control of major loads, selected devices (water heaters, heat pumps, EV charging) were automatically controlled. The dataset includes the corresponding price profiles and labels for each meter.

Previous publications using parts of the dataset include analyses of grid-level peak reductions achieved by direct load control \cite{KaiserResidentialPeak2026}{} and time-variable grid tariffs combined with automated load control \cite{KaiserPeakReduction2026}{}. An additional study focuses on the manual responses of opt-in and not-opt-in project participants to a time-of-use tariff \cite{KreftShowcasingVolunteer2026}{}, showcasing the impact of the recruiting strategy on pilot results.

\section{Methods}
The dataset consists of four main data types: 15-minute smart meter measurements, metadata on each meter and its corresponding installation, metadata on participation in the pilot project, and price profiles. The following subsections describe the methods related to each of these data types.

\subsection{Smart Meter Measurements}
The underlying dataset for this work comprises 15-minute cumulative energy readings since installation from 2,788 meters. The meters record both the active and the reactive energy withdrawn from and injected into the grid. However, reactive energy is recorded only during active energy withdrawal.
The corresponding OBIS codes\cite{VSEMeteringCode2025}{}, a standard to label different types of measurements, are included in Table~\ref{tab:smart_meter_data}.

\begin{table}
    \small
    \centering
    \begin{tabularx}{\textwidth}{lXl}
    \toprule
    Column header & Description & OBIS Code\\
    & & (cumulative)\\
    \midrule
    timestamp\_utc & Coordinated Universal Time (UTC) timestamp at the start of the 15-minute interval in the format \textit{YYYY-MM-DD hh:mm:ss+00:00}. & -\\
    kWh\_to\_installation & Active energy withdrawn from the grid. & 1-1:1.8.0*255\\
    kWh\_to\_grid & Active energy injected into the grid. & 1-1:2.8.0*255\\
    kvarh\_to\_installation & Reactive energy withdrawn from the grid during times of active energy withdrawal. & 1-1:5.8.0*255\\
    kvarh\_to\_grid & Reactive energy injected into the grid during times of active energy withdrawal. & 1-1:8.8.0*255\\
    \bottomrule
    \end{tabularx}
    \caption{Smart meter data description.}
    \label{tab:smart_meter_data}
\end{table}

Data from individual meters are collected via a smart metering system. In this case, two types of power line communication technology are used. A data concentrator, typically installed at the transformer station, collects data from the meters within its transformer service area and transmits them to the head‑end system in the backend. The original dataset consists of raw metering data that have not been subject to validation or correction. Consequently, data gaps and inconsistencies may occur.

To prepare the dataset for publication, the cumulative readings are first resampled to a 15-minute time series to ensure no time steps are missing. Then, the energy values for each 15-minute interval are computed as the difference between two readings, where available. While the majority of the data is consistent, there are rare instances where this calculation yields a negative value for at least one energy type (six instances across the final dataset). In these instances, the values for the respective meter and time step are set to \textit{not-a-number (NaN)}.

Smart meter profiles of buildings with a newly installed photovoltaic (PV) system are temporally truncated to prevent the possibility of mapping a given profile to a specific building based on its PV installation date.
A smart meter is assumed to monitor a PV system if the value for the active energy fed back to the grid exceeds 0.025 kWh (corresponding to a mean power of 100~W over 15~minutes) in at least 10 time steps, considering data from July 2022 to December 2024. The first time step exceeding this threshold is considered the installation date. Once identified, all data for the year of installation is removed to prevent disclosure of the installation date and ensure privacy. The maximum net injection in the remaining 2023/24 data is added to the metadata (cf. Table~\ref{tab:metadata}).

Finally, the dataset is filtered according to the exclusion criteria defined in Table~\ref{tab:exclusions}. Applying these criteria reduces the initial count of 2,788 meters to 2,447 meters. This subset represents the final group of smart meters included in this publication.
\begin{table}
    \centering
    \begin{tabularx}{\textwidth}{Xc}
    \toprule
    Criterion & \parbox[t]{2.2cm}{\centering Number of \\ smart meters} \\
    \midrule
    Missing installation or object identifier (see the next subsection) & 30 \\
    Non-residential object type & 179 \\
    Non-residential installation type & 24 \\
    Meter only shows the sum of other meters & 5 \\
    Inconsistent data (260 and 461 time steps with negative values) & 2 \\
    PV installation date revealed in another publication & 3 \\
    Unique features regarding the maximum injection or major loads that may raise privacy concerns & 33 \\
    Less than 12 weeks of continuous data in 2023/24 & 65 \\
    \bottomrule
    \end{tabularx}
    \caption{Criteria for excluding smart meters from the dataset. The criteria are applied in the given order, resulting in the stated number of exclusions per criterion.}
    \label{tab:exclusions}
\end{table}

\begin{figure}
    \centering
    \includegraphics[width=\textwidth]{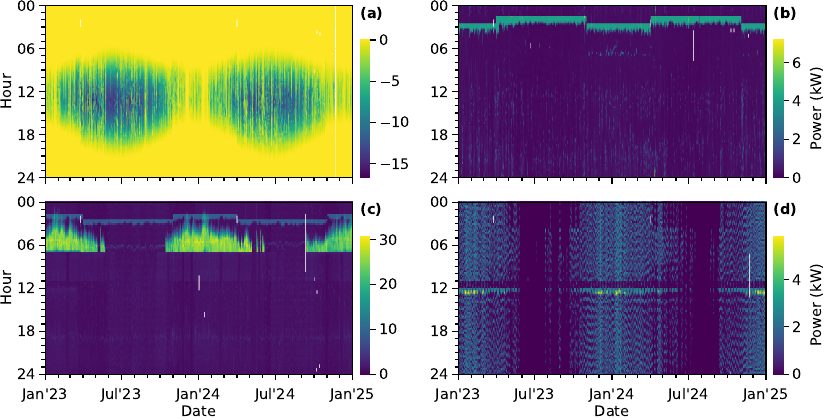}
    \caption{Four characteristic smart meter profiles: (a)~a feed-in system, (b)~a single-family house with an electric water heater which operates at the start of the unblocked time period during the night (different ripple control times apply from November to March and April to October), (c)~a single-family house with a storage heating system, and (d)~a separately monitored heat pump which is blocked from 11~am to noon on weekdays. The heatmaps show the 15-minute average active power in local time (Europe/Zurich).}
    \label{fig:example_profiles}
\end{figure}

Fig.~\ref{fig:example_profiles} visualizes the data for four meters with characteristic profiles. The heatmaps reveal distinct daily patterns of high injection and consumption. While the feed-in system's injection in Fig.~\ref{fig:example_profiles}(a) is driven by global radiation, the shown consumption patterns are mainly driven by ripple control, a widespread control strategy for electric water heaters and heating systems. The ripple control groups water heaters by storage size and power, and confines each group's operation to a specified period during the night. As visualized in Fig.~\ref{fig:example_profiles}(b), this results in recurring nightly peaks. Similarly, electric storage heaters are allowed to operate between 11~pm and 7~am, resulting in high consumption during these hours in winter (Fig.~\ref{fig:example_profiles}(c)). Furthermore, heat pumps can be blocked for up to four hours a day, leading to daily interruptions in their consumption pattern (Fig.~\ref{fig:example_profiles}(d)). In the given dataset, all electric water heaters, heat pumps, and storage heaters that are registered with the utility are controlled via ripple control.

\subsection{Smart Meter Metadata}
The metadata file consolidates data from multiple sources and provides key information for each smart meter, as described in Table~\ref{tab:metadata}. Each meter monitors a specific scope, referred to as an installation, comprising either an entire building (e.g., a single-family house), a building subarea (e.g., an apartment), or an individual sub-metered component (e.g., a heat pump or a feed-in system). The metadata file provides the given installation type and the major loads connected, including electric water heaters, electric heating systems, and EV charging stations. The device information relies on customers' and installers' reporting of equipment installation or removal. Although reporting is mandatory \cite{VSEWerkvorschriftenCH2025}{}, the records can be incomplete or outdated. Smart meter data suggest that particularly removals are not consistently reported.
To verify the actual presence of a specific device, the characteristic load profiles visualized in Fig.~\ref{fig:example_profiles} serve as an additional reference. For nine meters, a change in the device configuration was reported during the relevant period. For these, the column ``1\_change'' is set to \textit{True}, and all devices that were present at some point in time are indicated.

\begin{table}
    \small
    \centering
    \begin{tabularx}{\textwidth}{lX}
    \toprule
    Column header & Description \\
    \midrule
    0\_meter\_id & Unique smart meter identifier.\\
    0\_installation\_type & Installation type monitored by the smart meter, e.g., apartment, single-family house, heat pump, or common areas.\\
    0\_object\_type & Type of the object to which the installation belongs, e.g., single-family house, apartment building, or underground garage.\\
    0\_object\_id & Object identifier, included for (i) objects with one housing unit (installation of the type ``apartment'' or ``single-family house'') and (ii) non-housing installations of other objects.\\
    0\_object\_id\_manual & True means that the object ID was refined manually.\\
    0\_num\_housing\_units & Number of housing units belonging to the same object as the given non-housing installation, e.g., a heat pump.\\
    0\_num\_data\_points & Number of 15-minute time stamps with all energy values available.\\
    0\_max\_net\_injection & Maximum 15-minute average net injection into the grid in kW.\\
    1\_ewh & True indicates that the meter monitors an electric resistance water heater.\\
    1\_hp & True indicates that the meter monitors a heat pump.\\
    1\_hp-add & True indicates that the heat pump has a backup heating rod.\\
    1\_hp-wh & True indicates that the meter monitors a domestic hot water heat pump.\\
    1\_ev & True indicates that the meter monitors a charging station for electric vehicles.\\
    1\_storage\_heating & True indicates that the meter monitors a storage heating system.\\
    1\_direct\_heating & True indicates that the meter monitors a direct heating system.\\
    1\_change & True means that there has been a change in the device configuration between the first and the last data point of the given meter; all devices that were present at some time are labeled.\\
    2\_energy\_tariff & Energy tariff type during the pilot. None means that the standard tariff applies.\\
    2\_grid\_tariff & Grid tariff type during the pilot. None means that the standard tariff applies.\\
    2\_tariff\_from & Start date of the pilot tariff in the format \textit{YYYY-MM-DD} in local time.\\
    2\_tariff\_to & End date of the pilot tariff in the format \textit{YYYY-MM-DD} in local time (the given date is not included).\\
    2\_wh\_control & True means that the pilot load control device automatically controlled the electric water heater or domestic hot water heat pump (all meters of a given household are marked).\\
    2\_hp\_control & True means that the pilot load control device automatically controlled the heat pump (all meters of a given household are marked).\\
    2\_ev\_control & True means that at least one charging process of an electric vehicle owned by the customer corresponding to the given meter has been controlled during the pilot.\\
    \bottomrule
    \end{tabularx}
    \caption{Metadata description. Columns starting with ``0\_'' provide general information, columns with ``1\_'' indicate the presence of major loads, and columns with ``2\_'' refer to participation in the pilot project.}
    \label{tab:metadata}
\end{table}

Furthermore, each installation belongs to a specific object with a corresponding identifier and object type. However, a specific object identifier may be assigned to multiple installations. For example, an object of the type ``apartment building'' may cover several apartments, one heat pump, and one installation of the type ``common areas''.
The object types considered residential are (i)~detached single-family houses, two-family houses, and townhouses, which are referred to as ``single-family house'', (ii)~multi-family houses and housing complexes, which are referred to as ``apartment building'', and (iii)~underground garages, for which it was validated that they belong to residential buildings. As underground garages and other common installations may belong to several buildings, they may have their own object identifier and might not be mapped to all buildings that have access to them. For those with at least one electric water heater, heat pump, or EV charging station, we manually checked whether a mapping based on addresses is possible and assigned a common object identifier to all buildings that have access to the given underground garage or common installation, if applicable. Additionally, we resolved cases where the object information from two sources did not match. When the sources disagreed on whether two installations belonged to the same object or separate objects, we used satellite imagery to determine the correct mapping. Meters related to one of the described cases are labeled in the column ``0\_object\_id\_manual''.

Even though all identifiers are replaced by randomly assigned integer values, customers whose data are included in the dataset can identify themselves using their electricity data from bills or the utility's online customer portal. This may enable identifying a neighbor's profile via the common object identifier. To ensure anonymity, the object identifiers are only included for (i)~objects with only one housing unit (i.e., one installation of the type ``apartment'' or ``single-family house''), which allows mapping separately monitored heat pumps to the remaining household load, and (ii)~non-housing installations (e.g., heat pumps or common areas) of other objects. Nonetheless, to provide some context for these non-housing installations, the number of housing units belonging to the same object is specified.

\subsection{Pilot Participation}
Some customers associated with the given smart meters participated in a pilot project that evaluated three different demand response schemes for transformer-level peak reductions. All participants received a flat energy tariff instead of the standard two-tier energy tariff and were assigned to one of three new grid tariffs: (i)~an adjusted time-of-use (ToU) tariff with fixed seasonal tariff profiles, (ii)~a dynamic real-time tariff that can change every 15~minutes, or (iii)~a flat tariff. A best-accounting policy was in place to ensure that participants do not pay more than with the standard tariff. All participants could view their tariff on an online platform and manually shift consumption.

Additionally, 64 single-family houses with electric water heaters and/or heat pumps were equipped with a new load control device that replaced the standard ripple control and enabled individualized switching commands, and 42 customers registered their EVs for automatic control of home charging sessions.
In the ToU and real-time tariff setting, the automatic control aimed to minimize electricity cost by shifting consumption to low-price periods. In the direct load control setting with the flat grid tariff, the central system directly computed and communicated the switching commands for peak reduction. Electric water heaters and heat pumps without a pilot load control device, as well as storage heaters, remained under the standard ripple control. The installation date for the pilot load control device varies across households within the time window from September to December 2023.
The first controlled EV charging session occurred in June 2024.

The tariff and control assignment is included in the metadata file (cf. Table~\ref{tab:metadata}). The final dataset includes 606 meters associated with a pilot participant, 66 meters with the pilot control for the electric water heater or heat pump, and 18 meters belonging to customers with at least one controlled EV charging session.

\subsection{Price Profiles}
In Switzerland, customers with an annual electricity consumption of less than 100,000~kWh cannot (yet) choose their electricity supplier freely. Therefore, all customers relevant for the dataset purchase their electricity from the local utility Elektrizitätswerke des Kantons Zürich (EKZ), who determines both the energy and the grid tariff. The tariff data provided in the dataset include the price profiles for the standard EKZ energy and grid tariff, and the pilot tariffs mentioned above (cf. Table~\ref{tab:tariff_data}). The standard EKZ energy tariff in 2023/24 and the grid tariff in 2023 are two-tier ToU tariffs with higher prices from 7~am to 8~pm on weekdays. In 2024, the two-tier grid tariff was replaced by a flat tariff, reducing the overall price difference between the two tariff levels.

\begin{table}
    \small
    \centering
    \begin{tabularx}{\textwidth}{@{\hspace{5pt}}@{\extracolsep{\fill}}lX@{\hspace{5pt}}}
    \toprule
    Column header & Description\\
    \midrule
    timestamp\_utc & Coordinated Universal Time (UTC) timestamp at the start of the 15-minute interval in the format \textit{YYYY-MM-DD hh:mm:ss+00:00}.\\
    energy\_standard & Standard EKZ energy tariff (EKZ Mixstrom).\\
    energy\_unit\_low & Pilot energy tariff: Unit tariff based on the low tariff value in the standard tariff.\\
    energy\_unit\_cn & Pilot energy tariff: Unit tariff that ensured cost neutrality for inflexible consumers.\\
    grid\_standard & Standard EKZ grid tariff (EKZ Netz 400F).\\
    grid\_tou & Pilot grid tariff: Time-of-use price signal.\\
    grid\_dyn & Pilot grid tariff: Dynamic real-time price signal.\\
    grid\_unit & Pilot grid tariff: Unit tariff.\\
    \bottomrule
    \end{tabularx}
    \caption{Tariff data description. All values are in Rp./kWh excl. value added tax (VAT). The columns ``2\_energy\_tariff'' to ``2\_tariff\_to'' in Table~\ref{tab:metadata} specify for which meters and time periods the pilot tariffs apply.}
    \label{tab:tariff_data}
\end{table}

As the pilot focused on novel grid tariffs, participants received a flat energy tariff that would not interfere with the time-varying grid tariff component. Pilot participants with automatic control received the low-tariff component of the standard energy tariff as their energy tariff at all times. For pilot participants without automatic control, the standard energy tariff was converted to a flat tariff that ensured cost neutrality for inflexible consumers.
The pilot ToU grid tariff featured separate tariff profiles for summer and winter. From May to September, a low-price window from 1~pm to 6~pm and a mid-price window from midnight to 7~am applied. In the remaining months, the tariff profile featured a constant low-price period from midnight to 6~pm and a high-price period from 6~pm to midnight. In the initial pilot phase, the real-time price was proportional to the estimated inflexible load at the transformer station level. Starting from July 2024, a reinforcement learning agent computed the 15-minute price signal in real time. Finally, the flat grid tariff for the direct load control setting in 2023 was computed the same way as the cost-neutral energy tariff, while in 2024, the standard flat EKZ grid tariff applied. For details, we refer the reader to the related publications\cite{KaiserOrtsNetz2025, KaiserPeakReduction2026}{}.

\section{Data Records}
\begin{figure}
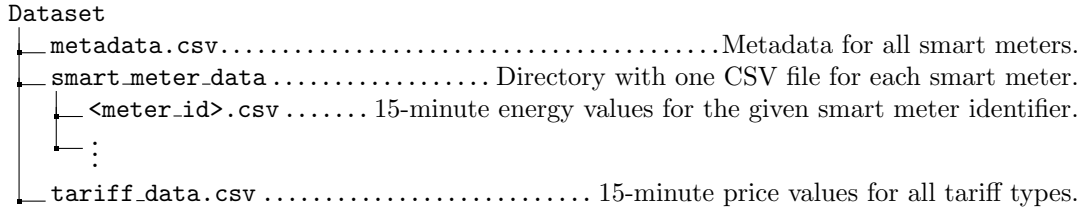

    \centering
    \begin{minipage}{\textwidth}
    \dirtree{%
    .1 Dataset.
    .2 {metadata.csv}\DTcomment{Metadata for all smart meters.}.
    .2 smart\_meter\_data\DTcomment{Directory with one CSV file for each smart meter.}.
    .3 {<meter\_id>.csv}\DTcomment{15-minute energy values for the given smart meter identifier.}.
    .3 \vdots.
    .2 {tariff\_data.csv}\DTcomment{15-minute price values for all tariff types.}.
    }
    \end{minipage}
    \caption{File structure.}
    \label{fig:filestructure}
\end{figure}

Fig.~\ref{fig:filestructure} gives the structure of the dataset, which is publicly available at Zenodo\cite{KaiserZenodoSM2026}{}.
It contains one CSV file with the metadata, a separate CSV file with the 15-minute energy data for each smart meter, and one CSV file with the 15-minute price values for all tariff types. Tables~\ref{tab:smart_meter_data} to \ref{tab:tariff_data} further describe each column in each file. The data cover the years 2023 and 2024, where available.

The dataset does not include weather data. However, the Swiss Federal Office for Meteorology and Climatology provides data at 10-minute resolution, which are freely accessible \cite{OpenDataMeteoswiss}{}. The closest weather station to the installations in this dataset is ``Zürich/Kloten''.

\section{Technical Validation}
This section gives insights into data availability and validates the data by comparing it to standard values and additional measurements.
\subsection{Smart Meter Data Availability}
\begin{figure}
    \centering
    \includegraphics[width=\textwidth]{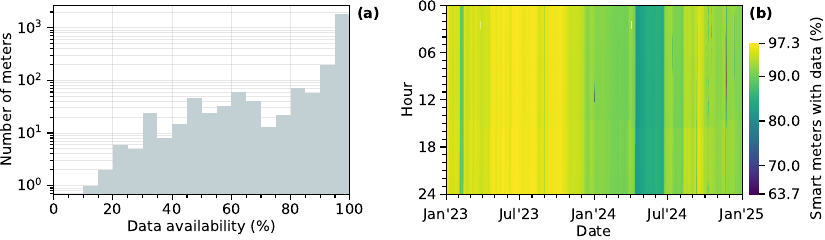}
    \caption{Data availability per smart meter (a) and percentage of smart meters with data over time (b).}
    \label{fig:available_data}
\end{figure}

Fig.~\ref{fig:available_data} visualizes the data availability per smart meter and the percentage of smart meters with data over time. In both cases, the data at a particular time step are only considered available if there are measurements for all energy values. Fig.~\ref{fig:available_data}(a) shows that for the majority of the smart meters (1,820 meters), there are measurements for all energy values for at least 95~\% of the time steps. Possible reasons for incomplete data for the remaining meters include removal of data due to a new PV installation (see Section Methods), interruptions caused by hardware replacements, or problems with power line communication.

\subsection{Annual Consumption}
\begin{figure}
    \centering
    \includegraphics[width=\textwidth]{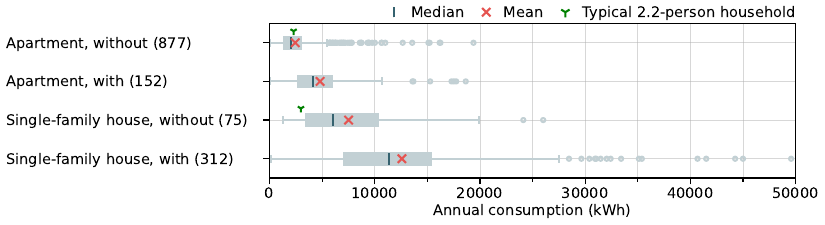}
    \caption{Average annual consumption for apartments and single-family houses. Separate distributions are shown for installations without an electric (water) heating system or an EV charging station, and installations with at least one of these devices according to the utility's records (cf.~Table~\ref{tab:metadata}). The values in parentheses correspond to the number of installations.}
    \label{fig:annual_consumption}
\end{figure}

Fig.~\ref{fig:annual_consumption} shows distributions of the average annual active energy consumption across 2023 and 2024 for apartments and single-family houses, distinguished by installations without an electric (water) heating system or an EV charging station and with either one of these. It includes data from installations with a data availability of 95~\% or more and no PV. Additionally, the figure shows the annual consumption of a typical household without electric (water) heating or an EV charging station. These reference values are based on the annual consumption values published in a fact sheet by the Swiss Federal Office of Energy \cite{SFOEStromverbrauchTypischerHaushalt2021} (2,190~kWh for a 2-person apartment $\pm$ 458.5~kWh per person and 4,048~kWh for a 4-person single-family house $\pm$ 593.5~kWh per person), and the average household size of 2.2 persons in the considered municipality \cite{DatasetDurchschnittlicheHaushaltsgrösse2025}{}. The value for apartments matches the mean and median of the households considered in the dataset. However, the single-family houses in the dataset exhibit a higher electricity consumption, possibly due to above-average household sizes and the presence of larger electric devices that are not registered with the utility. A manual screening of the corresponding smart meter profiles supports the latter assumption.

\subsection{Daily Energy Values and Weather Data}
\begin{figure}
    \centering
    \includegraphics[width=\textwidth]{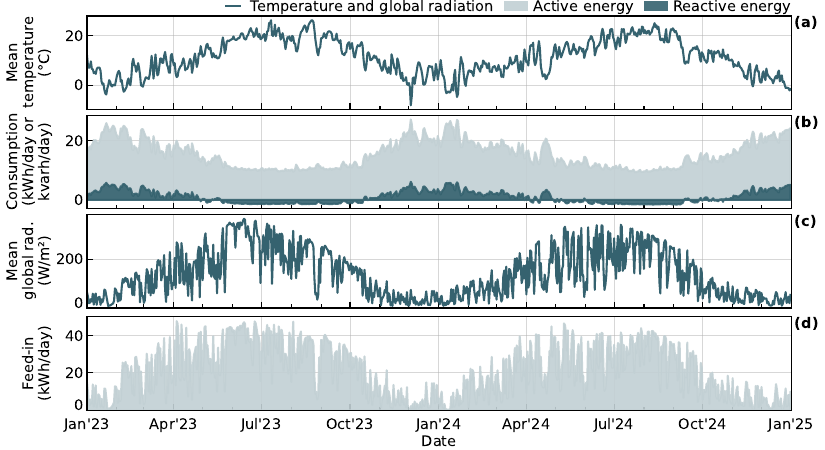}
    \caption{Daily weather and energy data: (a)~ambient temperature, (b)~active (light blue) and reactive (dark blue) net consumption of installations without PV, (c)~global radiation, and (d)~injection by PV feed-in systems. Weather parameters represent daily means, while energy values represent daily sums, averaged across the corresponding installations.}
    \label{fig:daily_values}
\end{figure}

Fig.~\ref{fig:daily_values} shows the daily mean temperature and global radiation\cite{OpenDataMeteoswiss} alongside the daily active and reactive energy consumption and active energy injection by PV. The net consumption values in Fig.~\ref{fig:daily_values}(b) represent the average across all installations with at least 95~\% data availability and no PV. The injection values in Fig.~\ref{fig:daily_values}(d) represent the average across all meters monitoring solely a feed-in system. The figure confirms that active energy consumption increases with decreasing temperature, driven by increased demand for electric heating. This seasonal trend is also reflected in reactive energy consumption, which increases during winter due to heat pump operation. The reactive power injection during summer in Fig.~\ref{fig:daily_values}(b) is consistent with the capacitive behavior observed, e.g., in a study by a Finnish distribution system operator\cite{PihkalaAnalysisChanging2019}{}. Furthermore, the figure validates that the recorded PV generation is consistent with global radiation data.

\section{Usage Notes}
All CSV files use semicolons as column separators. For analyzing the smart meter data of automatically controlled loads (see Section Methods), we recommend focusing on the periods listed in Table~\ref{tab:periods_ewh_hp}, as done in related publications\cite{KaiserOrtsNetz2025, KaiserPeakReduction2026, KaiserResidentialPeak2026}{}.
\begin{table}
    \footnotesize
    \centering
    \addtolength{\tabcolsep}{-0.03cm}
    \begin{tabularx}{\textwidth}{lllX}
    \toprule
    Scheme & Ripple control & No control & Pilot control\\
    \midrule
    Time-of-use tariff & 01.01.-14.09.2023 & 20.12.2023-16.01.2024 & 18.07.-04.09.2024 \& 19.11.-19.12.2024\\
    Dynamic real-time tariff & 01.01.-14.09.2023 & 20.12.2023-22.01.2024 & 18.07.-04.09.2024 \& 19.11.-19.12.2024\\
    Direct load control & 01.01.-14.09.2023 & 20.12.2023-27.02.2024 & 29.08.-04.09.2024 \& 19.11.-19.12.2024\\
    \bottomrule
    \end{tabularx}
    \caption{Relevant periods for automatically controlled electric water heaters and heat pumps.}
    \label{tab:periods_ewh_hp}
\end{table}

\section{Data Availability}
The smart meter measurements, metadata, and tariff data are available at Zenodo\cite{KaiserZenodoSM2026}{}. Weather data can be retrieved from MeteoSwiss \cite{OpenDataMeteoswiss}{}. The closest weather station to the installations in this dataset is ``Zürich/Kloten''. 

\section{Code Availability}
The code for analyzing the dataset was implemented in Python 3.10 and is available on GitHub (\url{https://github.com/ka-kai/smart-meter-data}).

\section{Author Contributions}
\textbf{K.K.}: Data curation, Software, Formal Analysis, Visualization, Writing - original draft, Writing - review \& editing. 
\textbf{M.K.}: Data curation, Software, Writing - review \& editing.
\textbf{M.G.V.}: Data curation, Supervision, Writing - review \& editing.
\textbf{G.H.}: Supervision, Writing - review \& editing.

\section{Acknowledgements}
The authors would like to thank all colleagues at EKZ who were involved in data collection and Roman Käslin from EKZ for his valuable input during the preparation of the dataset.

\section{Funding}
This work is part of a project that received funding from the Swiss Federal Office of Energy (SFOE) under grant number SI/502271.

\end{document}